\documentclass[11pt]{article}
\pdfoutput=1
\usepackage{moriond,epsfig}
\usepackage{overpic}

\def\Journal#1#2#3#4{{#1} {\bf #2}, #3 (#4)}

\def\PRD{{\em Phys. Rev.} D}

\def\be{\begin{equation}}
\def\ee{\end{equation}}
\def\bea{\begin{eqnarray}}
\def\eea{\end{eqnarray}}

\newcommand{\phiphi}{\ensuremath{B_s^0 \to \phi \phi}}
\newcommand{\jpsiphi}{\ensuremath{B_s^0 \to J/\psi \phi}}
\newcommand{\LLppp}{\ensuremath{\Lambda_b^0 \to \Lambda_c^+ \pi^- \pi^+ \pi^-}}
\newcommand{\dz}{D\ensuremath{\emptyset}}
\newcommand{\ifb}{\ensuremath{\mathrm{fb^{-1}}}}
\newcommand{\ipb}{\ensuremath{\mathrm{pb^{-1}}}}

\newcommand{\mevcc}{MeV/\ensuremath{c^2}}

\begin{document}
\vspace*{4cm}
\title{MEASUREMENTS OF THE MASSES, LIFETIMES AND DECAY MODES OF HADRONS AT TEVATRON}

\author{M. DORIGO on behalf of the CDF and \dz\ collaborations}

\address{Department of Physics, University of Trieste,\\
and INFN, Trieste Section, Trieste, Italy}

\maketitle
\abstracts{The Tevatron provides 1.96 TeV p$\bar{\textup{p}}$ collisions and allows for collection
of rich $b$-hadron samples to the two experiments CDF and \dz. 
The study of heavy flavor properties represents a fruitful opportunity to investigate 
the flavor sector of the Standard Model (SM)
and to look for hints of New Physics (NP). 
Here we report the first measurement of polarization
amplitudes in $B^0_s$ charmless decays, world leading results on $b$-hadron lifetimes, 
and measurements of several other properties of $b$-hadrons.}

\section{\phiphi\ polarization measurement}\label{sec:phiphi}
The \phiphi\ decay proceeds through a $b \to s\bar{s}s$ 
quark level process, whose dominant
diagram is the $b \to s$ penguin: it is potentially sensitive
to NP that could manifest itself through 
the presence of new particles in the penguin loop.
In addition, the \phiphi\ is a decay of a pseudo-scalar meson to two vector mesons 
whose differential decay rate is determined by three independent
amplitudes corresponding to different polarizations: 
one longitudinal ($A_0$) and two transverse, 
with spins parallel ($A_\parallel$) or perpendicular 
($A_\perp$) to each other.\footnote{The three 
polarization amplitudes are constrained
by the unitarity condition: $|A_0|^2+|A_\parallel|^2+|A_\perp|^2=1$.}
In the SM, $|A_0|^2 \gg |A_\parallel|^2+|A_\perp|^2$
is naively expected in $B$ decays to two light vector mesons.\cite{naive_pred}
This was experimentally confirmed 
by BaBar and Belle in tree-dominated transitions.\cite{BsRR} 
In contrast, it was found $|A_0|^2\simeq |A_\parallel|^2+|A_\perp|^2$ in 
$b \to s$ penguin decays.\cite{BsPhiK} 
In order to shed light on this ``polarization puzzle'' additional
experimental information is required and $b \to s$  penguins in charmless $B_s^0$ decays are a promising opportunity.
The \phiphi\ decay has been observed in its $K^+K^-K^+K^-$ final state for the first time by CDF in 2005 in 180 \ipb\ of integrated luminosity: 
8 events have been counted, and the branching ratio ($\mathcal{B} $) has been measured.\cite{BR1}  
Recently, CDF presents an updated analysis with 2.9 \ifb\ 
of data collected with the displaced track trigger.\cite{BR2}
The reconstructed signal events are $295\pm20\textup{\small (stat)}\pm12\textup{\small (syst)}$ and 
$\mathcal{B}=[2.40 \pm 0.21\textup{\small (stat)} \pm 0.86 \textup{\small (syst)}] \times 10^{-5}$,
which is consistent with the previous result and with the theoretical prediction.\cite{beneke}

We report the first polarization measurement of such decays using
the same data set of the branching fraction update.\cite{pol}
In this analysis, the untagged time-integrated decay 
rate as a function of three angular variables of the final state decay products
is considered. The polarization amplitudes are corrected for 
the expected lifetime difference for the $B^0_s$
mass eigenstates using the world average $B^0_s$ lifetime and width difference,\cite{PDG} and the 
tiny $\mathsf{CP}$ phase in $B^0_s$ mixing is neglected, as expected in the SM.\footnote{Anyway,
the effect related to a possible non vanishing 
$\mathsf{CP}$-violating phase in mixing 
at a level consistent with the current world average is included in the systematic treatment.}$^,$~\cite{PDG}
Thus, a fit to the reconstructed $B^0_s$ mass and to the decay product angular distributions is performed.
The approach is validated by performing a similar measurement using \jpsiphi\ decays, 
collected via the same trigger, and comparing the obtained results with the current experimental 
information on the polarization in such a decay.\cite{jpsiphi} 
The measured polarization amplitudes and the cosine of $\delta_\parallel=\arg(A_0^\star A_\parallel)$ are:
$|A_0|^2 =  0.348\pm 0.041\textup{\small (stat)} \pm 0.021\textup{\small (syst)}$,
$|A_\parallel|^2 =  0.287 \pm 0.043 \textup{\small (stat)} \pm 0.011\textup{\small (syst)}$,     
$|A_\perp|^2 =  0.365 \pm 0.044 \textup{\small (stat)} \pm  0.027\textup{\small (syst)}$ and    
$\cos\delta_\parallel  = -0.91^{+0.15}_{-0.13}\textup{\small (stat)}\pm0.09\textup{\small (syst)}$. 
This measurement indicates that the expected amplitudes hierarchy is disfavored
in this charmless $B^0_s$ decay, being $|A_0|^2 < |A_\parallel|^2+|A_\perp|^2$.
The plot in Fig.~\ref{fig:pol} (a) shows the estimates point for $f_0=|A_0|^2$ versus
$f_\parallel=|A_\parallel|^2$ compared with the prediction of different
theoretical models developed in the SM.\cite{beneke,pol_th}
\begin{figure}[!ht]
\begin{overpic}[width=0.5\textwidth]{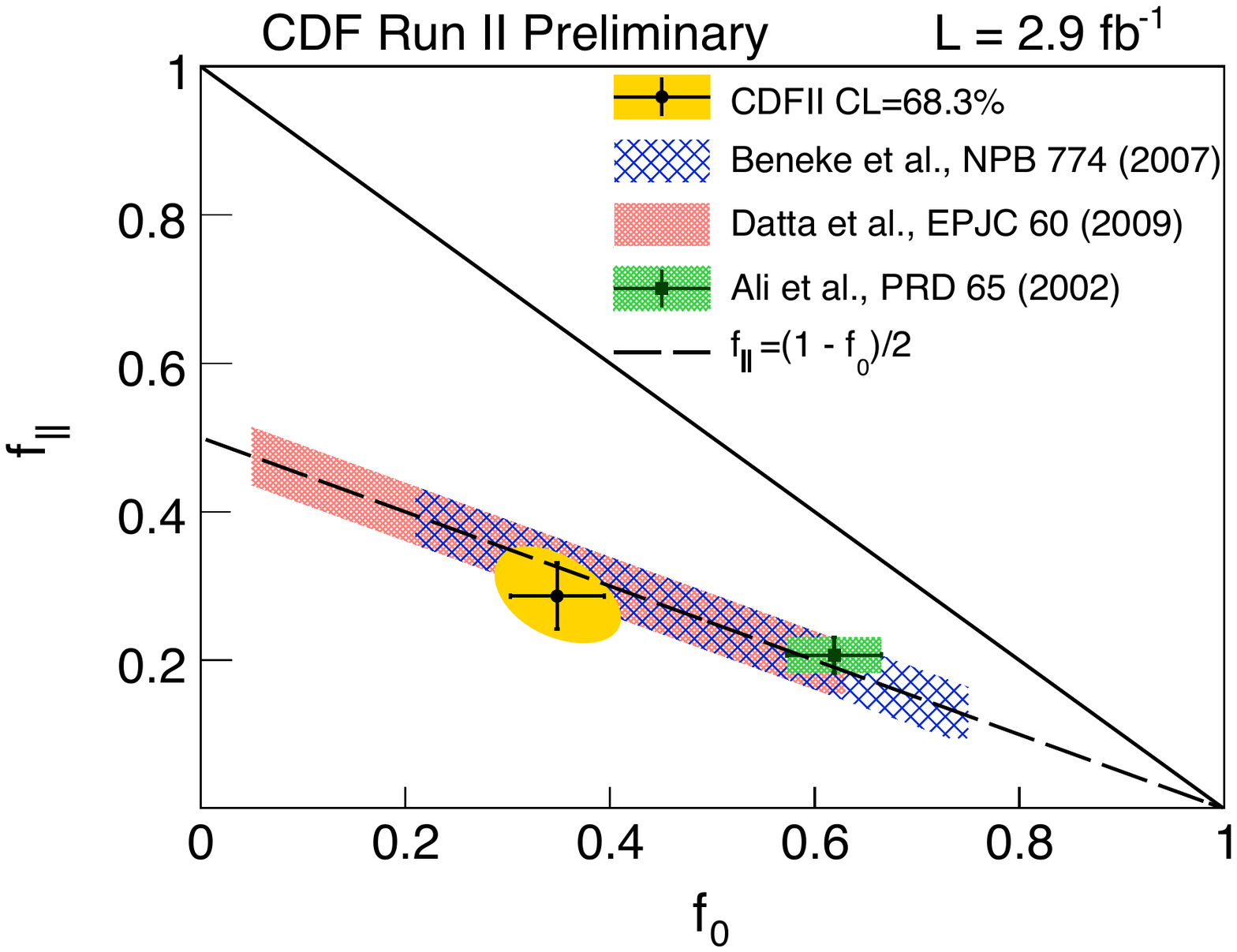}
\put(1,2){(a)}
\end{overpic}
\begin{overpic}[width=0.5\textwidth]{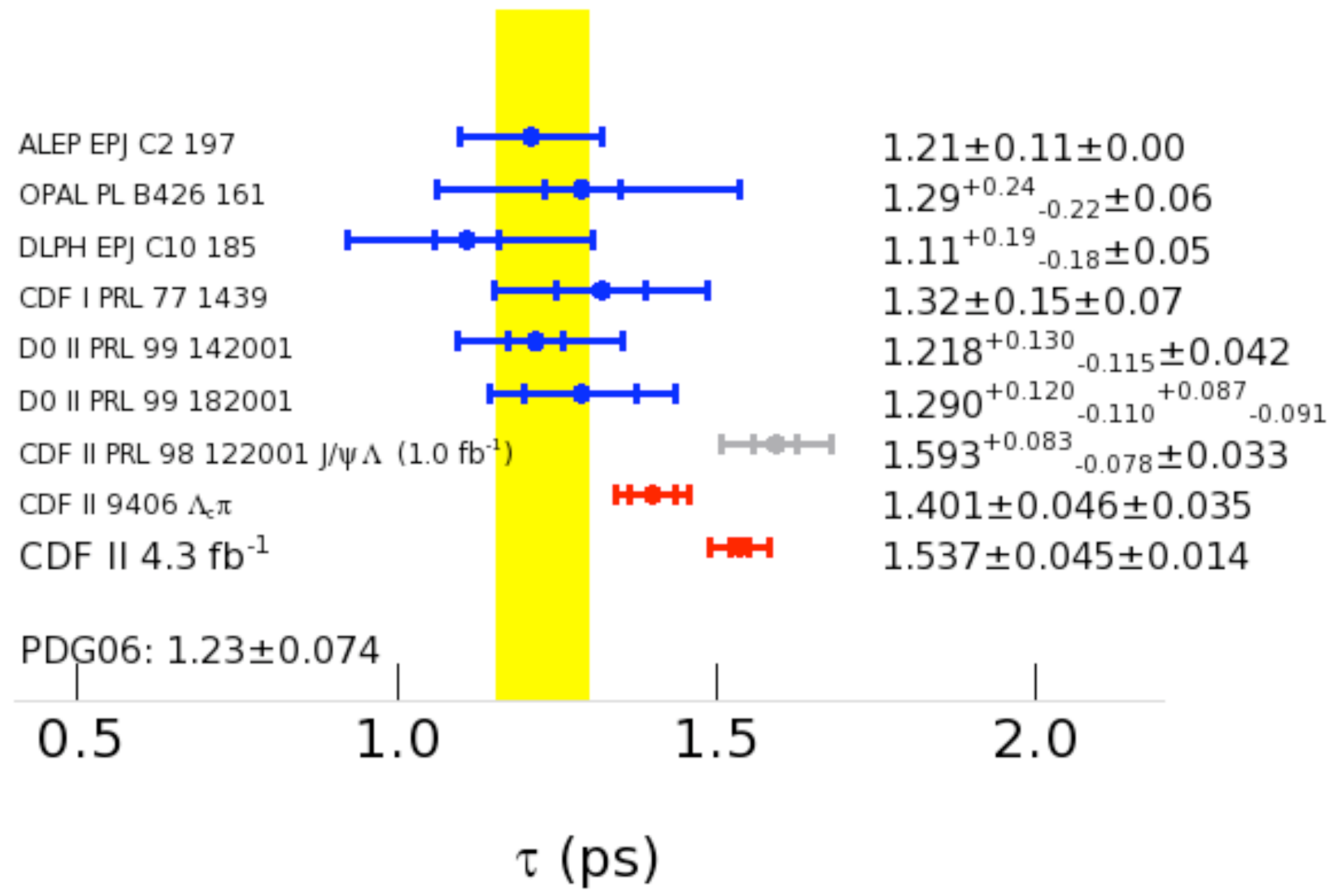}
\put(-1,2){(b)}
\end{overpic}
\caption{(a) Comparison between the \phiphi\ polarization measurement and several theoretical predictions. 
(b) $\Lambda_b^0$ lifetime: comparison between the new result and selected previous measurements.}
\label{fig:pol}
\end{figure} 

\section{Lifetimes}\label{sec:lifetimes}
The lifetime of the ground state hadrons containing a $b$ quark 
and lighter quarks is largely determined 
by the charged weak decay of the $b$ quark. 
Ignoring the lighter quarks in the hadron, the spectator model
predicts equal lifetimes for all $b$-hadrons.
However, several effects
change these expectations by up to about 10\%.
The lifetime ratio $\tau(B^+)/\tau(B^0)$ is predicted to be in the range 1.04--1.08;~\cite{B_life} 
for the $\tau(\Lambda_b^0)/\tau(B^0)$ the range is wider, 
from 0.83 to 0.93.\cite{L_life}

Using a 4.3 \ifb\ data sample, CDF searches for fully reconstructed $b$-hadron decays with a $J/\psi$ decaying 
to muon pairs: $B^+ \to J/\psi K^+$, $B^0 \to J/\psi K^{0\star}$, $B^0 \to J/\psi K^0_S$
and $\Lambda_b^0 \to J/\psi \Lambda^0$.~\cite{life_analysis} 
The data are collected by the dimuon trigger, which has no biasing
effects on the observed proper time distribution. 
The analysis consists of a maximum likelihood fit to the mass, 
the proper decay time and the proper decay time uncertainty of the reconstructed
candidates.
The measured lifetimes are reported in Tab.~\ref{tab:lifetimes}: they 
are the most precise determination of the $B^+$, $B^0$ and $\Lambda_b^0$ lifetimes.
Several systematic uncertainties have been studied with Monte Carlo samples; 
while the overall systematic uncertainties remain small, 
the uncertainty on the extracted lifetime values
is dominated by the silicon detector alignment uncertainty in
the case of $B$ mesons and by the resolution effects in the case of the $\Lambda_b^0$.
A cancellation of some common systematic uncertainties
in the lifetime ratios is achieved by using the vertex of the two tracks from the $J/\psi$, common to all decay modes,
 as an estimate of the transverse decay length.\footnote{The transverse decay length is
the projection of the decay lenght in the plane transverse to the beam.}

The $\Lambda_b^0$ lifetimes measurement has particular interest (see Fig.~\ref{fig:pol} (b)): until
2006 all measurements were in agreement and lay at the lower end of the theoretically expected
value. Then, CDF performed two high precision measurements \cite{life_CDF} which are significantly above previous
results; this is confirmed by the latest result here reported.
\begin{table}[!ht]\small
\begin{center}
\caption[Lifetime measurements of $b$-hadrons.]{Lifetime measurements of $b$-hadrons.}
\label{tab:lifetimes}
\medskip
\begin{tabular}{|cccc|}
\hline
     Hadron         & Lifetime  [ps] & Lifetime ratio (over $\tau(B^0)$) & PDG 08 [ps]\\
\hline
$B^+$         & $1.639 \pm 0.009 \textup{\small (stat)} \pm 0.009 \textup{\small (syst)}$   &  $1.088 \pm 0.009 \textup{\small (stat)} \pm 0.004 \textup{\small (syst)} $ & 
$1.638 \pm 0.011$\\
$B^0$         & $1.507 \pm 0.010 \textup{\small (stat)} \pm 0.008 \textup{\small (syst)}$   & 1 & $1.530 \pm 0.009$\\
$\Lambda_b^0$         & $1.537 \pm 0.045 \textup{\small (stat)} \pm 0.014 \textup{\small (syst)}$   &  $1.020\pm 0.030 \textup{\small (stat)} \pm 0.008 \textup{\small (syst)}$&
 $1.383^{+0.049}_{-0.048}$\\
\hline
\end{tabular}
\end{center}
\end{table}

\section{\LLppp resonance structure}\label{sec:Lambda}
Since CDF has recently observed the resonant structure in the 
$\Lambda_b^0 \to \Lambda_c^+\pi^-\pi^+l^-\nu$ decay mode,\cite{L_lep}
a similar resonance structure  is expected in the corresponding hadronic decay mode
\LLppp, where the $l^-\nu$ pair is replaced by a $ud$ quarks pair.

Last year CDF performed an analysis aimed at the first observation of the 
\LLppp\ and of the following intermediate resonant states: 
$\Lambda_c(2595)^+ \pi^-$, 
$\Lambda_c(2625)^+ \pi^-$, $\Sigma_c(2455)^{++} \pi^-\pi^-$ and
$\Sigma_c(2455)^0 \pi^+\pi^-$. 
The analysis uses an integrated luminosity of 2.4 \ifb\ of data collected by the CDF trigger on two displaced tracks.
The $\Lambda_c^+$ is reconstructed in the $p K^- \pi^+$ decay mode and three tracks, 
assumed to be pions, are added to reconstruct the \LLppp. 
The relative branching fractions of the resonant states 
to the total ($\mathcal{B}_{\textup{\small rel}}$) are measured 
using the yields of each decay mode estimated by fitting 
the data mass distributions. 
The final results are listed in Tab.~\ref{tab:lambda}.
\begin{table}[!ht]\small
\begin{center}
\caption[\LLppp\ resonance structure: yields and branching fractions.]
{\LLppp\ resonance structure: yields and branching fractions.}
\label{tab:lambda}
\medskip
\begin{tabular}{|lcc|}
\hline
     $\Lambda_b^0$ decay mode    & Yield & $\mathcal{B}_{\textup{\small rel}}$ in $10^{-2}$\\
\hline
$\Lambda_b^0 \to \Lambda_c(2595)^+ \pi^- $  & 
$46.6 \pm 9.7 \textup{\small (stat)}$   &  
$2.5 \pm 0.6 \textup{\small (stat)} \pm 0.5 \textup{\small (syst)}$ \\

$\Lambda_b^0 \to \Lambda_c(2625)^+ \pi^- $  & 
$114 \pm 13 \textup{\small (stat)}$ &
$6.2 \pm 1.0 \textup{\small (stat)} ^{+1.0}_{-0.9} \textup{\small (syst)}$  \\

$\Lambda_b^0 \to \Sigma_c(2455)^{++} \pi^-\pi^-$   & 
$81 \pm 15 \textup{\small (stat)}$   & 
 $5.2\pm 1.1 \textup{\small (stat)} \pm 0.8 \textup{\small (syst)}$ \\

$\Lambda_b^0 \to \Sigma_c(2455)^0 \pi^+\pi^-$   & 
$41.5 \pm 9.3 \textup{\small (stat)}$   &  
$8.9\pm 2.1 \textup{\small (stat)} ^{+1.2}_{-1.0} \textup{\small (syst)}$\\
\hline
\end{tabular}
\end{center}
\end{table}

\section{Masses: the $\Omega_b^-$ observation}\label{sec:masses}
The quark model predicts a rich spectrum of baryons containing
$b$ quarks.\cite{q_m_b_predic} In 2007, the accumulation of large data sets from the Tevatron 
allowed the first observation of new baryons, the $\Xi_b^-$ and the $\Sigma_b^{(\star)}$.\cite{xi_b,sigma_b} 
The $\Omega_b^-$ is the latest observed of such heavy states:
in 2008, \dz\ made its discovery using 1.3 \ifb\ of data,\cite{omega_b_D0}
while CDF observed it last year in 4.2~\ifb.~\cite{omega_b_CDF}
In both cases, the $\Omega_b^-$ observation is made through the decay chain 
$\Omega_b^- \to J/\psi \Omega^-$, where $J/\psi \to \mu^+\mu^-$, 
$\Omega^- \to \Lambda K^-$, and $\Lambda \to p \pi^-$.\footnote{Charge conjugate modes are included implicitly.} 
However, the two experiments measure a $\Omega_b^-$ mass in significant disagreement.
The \dz\ analysis is built on the $\Xi_b^-$ discovery; \cite{xi_b} 
a yield of $17.8\pm 4.9 \textup{\small (stat)} \pm 0.8 \textup{\small (syst)}$  $\Omega_b^-$
events is extracted,
with a significance of 5.4 Gaussian standard deviations 
that the observed peak is not due to background fluctuations.
The estimated $\Omega^-_b$ mass is $6165\pm 10\textup{\small (stat)}\pm 13\textup{\small (syst)}$~\mevcc.

In the $\Omega^-_b$ observation, in addition to
its mass, CDF measures for the first time its lifetime. Moreover, in the same analysis, CDF updates 
the $\Xi_b^-$ mass measurement and performs its first lifetime measurement.\footnote{Using 
the decay chain $\Xi_b^- \to J/\psi \Xi^-$, where $J/\psi \to \mu^+\mu^-$, 
$\Xi^- \to \Lambda \pi^-$, and $\Lambda \to p \pi^-$.}
CDF measures the $\Omega^-_b$ mass, $m=6054.4\pm 6.8\textup{\small (stat)} \pm 0.9\textup{\small (syst)}$~\mevcc,
and lifetime, $\tau=1.13 ^{+0.53}_{-0.40} \textup{\small (stat)} \pm0.02\textup{\small (syst)}$~ps,
using a signal of $16^{+6}_{-4}\textup{\small (stat)}$ (5.5 $\sigma$ significance), and the
$\Xi^-_b$ mass, $m=5790.9\pm 2.6\textup{\small (stat)} \pm 0.8\textup{\small (syst)}$~\mevcc, and
lifetime, $\tau=1.56 ^{+0.27}_{-0.25} \textup{\small (stat)} \pm0.02\textup{\small (syst)}$~ps.
The small systematic uncertainties in the CDF measurements are due to the ability 
to reconstruct the actual trajectory of the long-lived hyperons from their hits in the silicon tracker.

The disagreement between the CDF and \dz\ $\Omega_b^-$ mass measurements consists of about six standard deviations. 
In addition, the measured $\Omega_b^-$ production rates relative to the $\Xi_b$
are different between the two experiments, being
$f_{\textup{\tiny CDF}}=0.27 \pm0.12 \textup{\small (stat)} \pm 0.01\textup{\small (syst)}$ 
and $f_{\textup{\tiny \dz}}=0.80 \pm 0.32 \textup{\small (stat)} ^{+0.14}_{-0.22}\textup{\small (syst)}$.
Neither measurement is very precise; nevertheless, CDF indicates a rate substantially lower than \dz.
The mass obtained by CDF agrees with theoretical estimates.\cite{q_m_b_predic}
Clearly, further studies are needed to resolve the discrepancies 
and analysis updates are ongoing with the addition of new available data.

\section{Conclusions}
In the latest years, the CDF and \dz\ heavy flavor programs reached maturation,
yielding results that are competitive to the $B$ factories ones for the $B^\pm$ and $B^0$ properties measurement, and
complementary to them for the study of the $b$-baryons and the $B^0_s$ meson.
We presented here a small sampling of recent results, including the first measurement
of decay-polarization structure in a charmless $B_s^0$ decay, world-leading measurements of $b$-hadron lifetimes,
the structure study of a $\Lambda^0_b$ hadronic decay mode and the observation of the $\Omega_b^-$ baryon.
These are obtained using just a fraction of the presently available data-samples, which keep increasing at a pace of 70 \ipb\ per week. 
The large amount of data, and ever improving analysis technique suggest a few years of exciting competition with the LHCb experiment
that has just started its operations.

\end{document}



